\documentclass[twocolumn,showpacs,preprintnumbers,amsmath,amssymb]{revtex4}
\usepackage{graphicx}

\begin{document}
\title{Feasible scheme for measuring
experimentally the speed of the response of quantum states to the
change of the boundary condition}
\author{Guang Ping He}
 \email{hegp@mail.sysu.edu.cn}
\affiliation{School of Physics \& Engineering and Advanced Research
Center, Sun Yat-sen University, Guangzhou 510275, China}

\begin{abstract}
When the boundary condition of a quantum system changes, how fast
will it affect the state of the system? Here we show that if the
response takes place immediately, then it can allow superluminal
signal transfer. Else if the response propagates in space with a
finite speed, then it could give a simple explanation why our world
shows classicality on the macroscopic scale. Furthermore,
determining the exact value of this speed can either clarify the
doubts on static experiments for testing Bell's inequality, or
support the pilot-wave interpretation of quantum mechanics. We
propose an experimental scheme for measuring this speed, which can
be implemented with state-of-art technology, e.g., single-electron
biprism interference.
\end{abstract}

\maketitle

\newpage

\section{Introduction}

Quantum mechanics achieved great success in the past century and was proven
correct in almost any physical process on any scale. But due to some
anti-intuitional features of quantum mechanics, people keep wondering why
our world has to be quantum. Some even doubted the completeness of quantum
mechanics. Many quantum interpretation theories thus arose, e.g., the
Copenhagen interpretation, statistical interpretation, pilot-wave
interpretation, and many worlds interpretation, etc.. But many details of
these theories generally involve some quantities which are not the
observables of quantum mechanics, therefore hardly any existing experiment
can prove or disprove these interpretations.

Nevertheless, here we will propose an experimental scheme that can provide
some clues to the details of how quantum mechanics works. Thus it may serve
as a starting point for picking the correct quantum interpretation. We
consider the following problem. It is well-known that once the Hamiltonian
and boundary condition of a quantum system are provided, the state of the
system and its time evolution are completely determined by Schr\"{o}dinger
equation (or Klein-Gordon/Dirac equations in the relativistic case). But how
fast is the state determined by these elements? More specifically, when the
boundary conditions change, how fast will the wavefunction in another
location of space be affected?

It is important to note that what we consider here is different from
the existing results obtained from systems with time-dependent
Hamiltonian and/or fast changing boundary conditions (e.g., Refs.
\cite{ft35,ft34}), where it was assumed that at any given time $t$,
the wavefunction $\psi (t)$ in the whole space satisfies the
Schr\"{o}dinger equation of the same $t$. In literature, this
assumption was widely adopted in quantum mechanics, despite that it
was not clearly stated as an assumption most of the time.
But it could violate the theory of relativity even if we replace Schr\"{o}%
dinger equation with Klein-Gordon or Dirac equation. In this paper,
instead of adopting this assumption, we are interested in whether
the Hamiltonian/boundary condition at a given time will affect the
wavefunction at all the locations of the space instantaneously, and
if the answer is no, then how fast the effect will occur. It is
worth noting that the problem cannot be solved only within the
framework of basic quantum mechanical formalism (e.g.,
Schrodinger/Klein-Gordon/Dirac equations). It has to rely on a
certain interpretation of the quantum theory to supply details on
how the quantum system \textquotedblleft gets
information\textquotedblright\ on the boundary condition.

We will show below that it is possible to measure experimentally the
speed how fast the response to the change of boundary condition will
take place. We will also show theoretically that if the speed is
infinite, then it can allow superluminal signal transfer and thus
conflicts with the theory of relativity. Else if the speed is
finite, then it may provide a clue to the long-time puzzling open
problem why our world looks classical on the macroscopic scale
despite that all microscopic processes are quantum. Furthermore,
knowing the exact value of this speed can help us to understand how
a quantum system \textquotedblleft knows\textquotedblright\ the
status of the boundary condition, which can help to judge whether
the pilot-wave theory or other quantum interpretations seems more
appropriate, and prove or disprove the doubts on static experiments
for testing Bell's inequality. Therefore, implementing our scheme
can provide results which will develop our understanding on quantum
mechanics, and bring us closer to the answer of John Wheeler's big
question \textquotedblleft why the quantum\textquotedblright\
\cite{qi256}.

\section{Importance of the speed}

Before going to the details of our experimental scheme, let us first
consider the Gedanken experiment illustrated in Fig. 1, to see why it is
important to determine the speed $v$ with which a quantum system responds to
the change of the boundary condition. A particle source produces single mode
quantum particles with speed $v_{0}$. The quantum particles can be either
photons, electrons, or neutrons, etc., and $v_{0}$\ can be either equal to
or smaller than the speed of light $c$. The output power of the source is
carefully controlled so that it produces only one particle at a time.
Similar to the \textquotedblleft which-way\textquotedblright\ experiment
\cite{qi735}, let the particles pass a double-slit wall (in fact we use
pinholes instead of slits) and then reach the screen. There is a barrier
behind pinhole A that can choose to either open or close the pinhole. The
status of pinholes A and B thus serves as the boundary condition for the
quantum state in the space. Let pinhole B be opened all the time. According
to quantum mechanics, when pinhole A is opened, a double-slit interference
pattern should be observed on the screen. On the other hand, if pinhole A is
closed, the interference pattern will disappear, while only the single-slit
diffraction caused by pinhole B will be observed. Now consider the following
question: suppose that pinhole A was initially opened, and is closed at time
$t_{1}$, then how fast will the pattern observed on the screen show a
response to this change of the status of pinhole A?

\begin{figure}[tbp]
\includegraphics{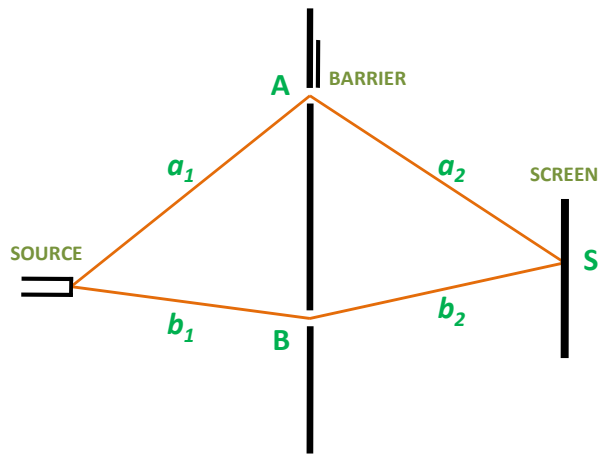}
\caption{Diagram of the double-slit experiment. Quantum particles
with speed $v_{0}$ are generated by the source, then pass the
pinholes A and B and are detected on the screen. Pinhole A can be
either opened or closed by the barrier behind it.}
\label{fig:epsart}
\end{figure}

On one hand, if the response takes place immediately, then it seems to allow
superluminal signal transfer and conflict with the theory of relativity.
This is because an observer at a certain location $S$ of the screen can
deduce whether a distant pinhole A is closed or not by the pattern he
observed. For simplicity, let $S$ be a point corresponding to a dark fringe
of the interference pattern when both pinholes A and B are opened. Suppose
that the particle source was initially shut down but turned on right after $%
t_{1}$. The observer then waits for a finite time interval $\Delta t$ ($%
\Delta t>(b_{1}+b_{2})/v_{0}$) so that a sufficient number of particles can
reach the screen from the source. Now if the observer detected a certain
amount of particle flux on point $S$ between the time $%
t_{1}+(b_{1}+b_{2})/v_{0}$ and $t_{1}+\Delta t$, he can deduce that pinhole
A was closed. Else, if he found that point $S$ is still dark after $%
t_{1}+\Delta t$, then he concludes that pinhole A was not closed at $t_{1}$.
Therefore, if the distance $a_{2}$ is sufficiently larger than $%
(b_{1}+b_{2})c/v_{0}$ so that $a_{2}>c\Delta t$, a superluminal signal is
transferred from pinhole A to point $S$. Although in practice the
interference will be too weak to detect if $a_{1}$\ and $a_{2}$ are very
large, in principle it still makes a difference in the observed pattern.
More importantly, unlike the spooky action at a distance realized with
Einstein-Podolsky-Rosen entangled pairs \cite{qi751} where the observers
cannot predict the outcome before performing the measurement, in our case
the difference in the pattern can indeed deliver a preassigned signal.
Therefore this result seems impossible according to the theory of relativity.

On the other hand, suppose that it takes a finite time before it
becomes possible to observe the response to the boundary condition
corresponding to the change of the status of pinhole A. That is, the
response propagates in space with a finite speed $v$. Then it is
puzzling what is traveling through space delivering the information
of the boundary condition. At the first glance, it seems to be a
natural interpretation that these traveling objects are the
particles themselves, who reach the boundary so that they
\textquotedblleft know\textquotedblright\ how many pinholes are
opened. But previous experiments showed that even if the particle
source produces only one single particle at a time, the interference
pattern will still present if both pinholes are opened. Then it may
seem weird to assume that a single particle reaches both pinholes
simultaneously by itself. But no matter the particle reaches a
pinhole in its entirety or by parts, as long as this interpretation
is correct, it seems natural that we should find $v=v_{0}$ (within
the precision allowed by the uncertainty principle). This is what
the mainstream interpretation theories of quantum mechanics predict.
But besides this picture, there are also other interpretations on
how quantum systems \textquotedblleft know\textquotedblright\ the
boundary condition. For instance, in literature there were doubts on
whether static experiments for testing Bell's inequality can provide
a convincing conclusion, because \textquotedblleft the settings of
the instruments are made sufficiently in advance to allow them to
reach some mutual rapport by exchange of signals with velocity less
than or equal to that of light\textquotedblright\ \cite{ft29,ft28}.
If this is also the case in our experiment, then it is possible that
the particle source, the slits, and the screen somehow
\textquotedblleft know\textquotedblright\ the status of each other
with or without the existence of the particle. Thus $v$ will not
have to be equal to $v_{0}$. According to the de Broglie-Bohm
pilot-wave interpretation of quantum mechanics
\cite{qi729,qi730,qi476}, the traveling objects carrying the
information of the boundary condition can be viewed as the pilot
waves set up in space by the Hamiltonian and boundary condition, and
the particle travels by following a certain pilot wave. In this
picture, both $v=v_{0}$\ and $v\neq v_{0}$\ are allowed in theory
\cite{qi729}. Specifically, if $v=c$, it may indicate that the
traveling objects are the virtual photons being exchanged between
the particle (or the source) and the boundary so that they
\textquotedblleft know\textquotedblright\ the existence of each
other, as described in quantum field theory. Besides these
interpretations, there can even be other mechanism that we may
currently be unaware of. Thus it is natural to assume in general
that $v_{0}\leq v\leq c$.

Obviously, if the value of $v$ can be measured, it can help us understand
the mechanism better. Furthermore, as long as $v$ is finite, we can have a
simple interpretation on the classicality of our macroscopic world. Consider
the following case. Suppose that the source was initially turned off, then
turned on and off intermittently after $t_{1}$. At each interval, it is
turned on only for a short period of time so that it produces only one
particle at the most. Let $\varepsilon $ denote the time between each
interval it is turned on. $\varepsilon $\ should be sufficiently long to
guarantee that the particle produced in the previous interval already
reached the screen and completely interacted with and was absorbed by the
screen, before the source is turned on in the next interval. Now if pinhole
A was initially closed, and the observer at point $S$ does not know whether
pinhole A is opened or closed at $t_{1}$, what pattern will he find on the
screen?

Since we assumed that the response to the change of the status of pinhole A
propagates in space with a finite speed $v$, if pinhole A is opened at $t_{1}
$, point $S$ will not be affected immediately. Therefore, if pinhole A is
sufficiently far away from the source and the screen, the interference
pattern should not be observed at point $S$ for a period of time, because
pinhole A is initially closed. But will the interference pattern be observed
later? If yes, then what makes the forthcoming particle interfere? Note that
it is assumed that the source produces only one particle at a time, and the
previous particle was already absorbed by the screen long before the next
particle is produced. Thus anything (for conciseness in the description, we
call it as pilot wave hereafter, no matter what it really is) generated by
the previous particle (if any) should no longer exist when the next particle
comes out. So if we assume that the pilot wave delivering the status of the
boundary condition is generated by the particle itself, then there is
nothing left from the previous particle to guide the next one. The next
particle has to generate its own pilot wave to sense the boundary.
Therefore, similar to the previous particles, the forthcoming particles will
not form an interference pattern at point $S$ either, as long as pinhole A
is so far away from the source and the screen, that each particle already
reached and was absorbed by the screen from the source before its own pilot
wave can reach the screen from pinhole A. Consequently, interference can
never be observed in this Gedanken experiment. If this is indeed the case,
then it gives a clue on why our world looks classical on the macroscopic
scale even though every single particle is ruled by quantum mechanics and
has wave-particle duality -- simply because in the real world most quantum
particles are closely surrounded by and interacting with other particles on
the microscopic scale, and they act as pinhole B and the screen to each
other, so that other boundary condition at a distance (which does not have
to be really far away on the macroscopic sense) is relatively too far to
have these quantum particles fully display their corresponding wave-like
behaviors such as interference. Of course, to fully explain the classicality
of our macroscopic world, we need to further study whether this mechanism
also plays a crucial role in every other complicated physical process. Thus
it is still too early to make a deterministic conclusion. Nevertheless, we
would like to pinpoint out that this interpretation has the advantage that
it does not require any new physical postulation. As long as $v$ is finite,
the above mechanism is valid, while $v$ has to be finite as long as
superluminal signal transfer is impossible. That is, this interpretation is
based merely on the validity of Special Relativity. Note that other existing
interpretations on the classicality of the macroscopic world generally
involve new postulations which may not have been proven. Therefore our
interpretation looks promising and worth further investigation.

On the contrary, if the interference pattern can indeed be observed in this
Gedanken experiment, then it seems to suggest that the pilot wave is
generated by the boundary instead of the particle itself. If so, then its
speed could be independent of the type of the particle. This picture is also
interesting since the corresponding mechanism will be worth studying, and it
may even be related to the interpretation of space-time structure and
gravity.

Either way, we can see that the corresponding physical picture is
interesting. Intuitively, it seems very possible that the result would be $%
v=v_{0}$. However, as mentioned in the Introduction, this speed cannot be
calculated from basic quantum mechanical formulas (e.g., Schr\"{o}dinger
equation) without involving any quantum interpretation theory. Therefore,
nothing should be taken for granted unless it is proven by experiments. Here
we are not going to reach a conclusion theoretically. Instead, we will
propose a feasible scheme to measure the speed $v$.

\section{The experimental scheme}

The experimental apparatus for measuring this speed is illustrated in Fig.
2. It is similar to the above Gedanken experiment, but none of the pinholes
needs to be far away so that it is practical to be implemented. There are
two wheels rotating clockwise (when viewing from the particle source) along
the same axis with the same speed $\omega $. The shape of the front wheel is
a sector of angle $\alpha $. It is located right behind pinhole A so it will
cover pinhole A from time to time as it rotates. The shape of the rear wheel
is a sector of angle $\beta $. It is located right in front of the screen so
it will cover a certain area of the screen as it rotates. Now suppose that
the speed $\omega $ is very high, and the output power of the particle
source is carefully controlled so that it produces no more than one single
particle at a time. In one round of the rotation of the wheels, at time $%
t_{1}$ the left edge $OE_{1}$ of the front wheel meets pinhole A so that A
is going to be covered by the front wheel (see Fig. 3a). At a later time $%
t_{2}$, the right edge $OE_{2}$ of the front wheel starts to leave pinhole A
so that A will be opened hereafter (see Fig. 3b). Let $OE_{0}$ denote the
line on the screen corresponding to the position of the left edge of the
rear wheel at $t_{1}$, and $OE_{3}$ denote the line on the screen
corresponding to the position of the right edge of the rear wheel at $t_{2}$%
. Then $OE_{0}E_{3}$ forms a sector with angle $\gamma =\beta -\alpha $
(supposing that the diameter of the pinhole is negligible).

\begin{figure}[tbp]
\includegraphics{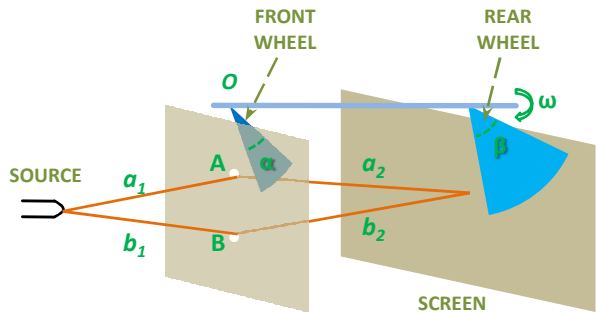}
\caption{The apparatus for measuring the speed of the response. Two
wheels rotating along the same axis with the same speed
$\protect\omega $ are placed behind pinhole A and in front of the
screen respectively, so that pinhole A and certain regions of the
screen are covered from time to time.} \label{fig:epsart}
\end{figure}

Now consider the pattern we will observe within the sector area $OE_{0}E_{3}$
of the screen after a sufficiently long period of time during which the
wheels completed many rounds of rotation. Note that in each round, the area $%
OE_{0}E_{3}$ of the screen is completed covered during the period from $%
t_{1} $ to $t_{2}$. On the other hand, pinhole A is kept opened before $%
t_{1} $ and after $t_{2}$. That is, the part of the screen within the area $%
OE_{0}E_{3}$ is exposed only when both pinholes A and B are opened.
Therefore, if the change of the boundary condition (i.e., the status of
pinhole A in our case) takes effect instantaneously (i.e., the response to
the change propagates in space with an infinite speed), only the
interference pattern will be observed within the area $OE_{0}E_{3}$.
Meanwhile, a mix pattern will be observed on the screen outside the area $%
OE_{0}E_{3}$, which is not only the interference pattern obtained when both
pinholes A and B are opened, but also overlaps with the single-slit
diffraction pattern when pinhole A is covered by the front wheel.

On the contrary, if the response has a finite speed $v$, then the mix
pattern will also be observed in some parts of the area $OE_{0}E_{3}$. This
is because pinhole A is covered during $t_{1}$ to $t_{2}$. And though it is
opened at $t_{2}$, the screen will not be affected until $t_{2}+\Delta T$,
where $\Delta T=a_{2}/v$. Therefore, if any particle reaches the screen
during $\Delta T$, it will contribute to form the single-slit diffraction
pattern as if pinhole A was not opened yet. Since the rear wheel is rotating
with the speed $\omega $, during $\Delta T$ its right edge sweeps through an
angle%
\begin{equation}
\delta =\omega \Delta T=\omega a_{2}/v,
\end{equation}%
thus leaving a narrow sector of angle $\delta $ at the left of $OE_{3}$
uncovered. Consequently, the mix pattern should be presented in this sector
area. Similarly, though pinhole A starts to close at $t_{1}$, it will not
take effect on the screen until $t_{1}+\Delta T$. Therefore, the
interference pattern instead of the mix pattern will be observed in a narrow
sector of angle $\delta $ to the left of $OE_{0}$. As a whole, there is
still a sector area of the screen in which only the pure interference
pattern will be observed. The sector has the same angle as that of the
sector $OE_{0}E_{3}$, but its position is like rotating the sector $%
OE_{0}E_{3}$ clockwise along the axis of the wheels by angle $\delta $,
while the pattern observed inside the sector (the position of the fringes)
does not rotate since the relative position of the pinholes and the screen
stays unvaried.

To observe the changed angle, in the experiment we can initially rotate the
wheels at an extremely slow speed $\omega _{s}$, while putting a
photographic plate on the screen and exposing for a long period of time
(just a little longer than that in an ordinary double-slit experiment
without the wheels), so that there can be a sufficient amount of particles
reaching the screen to form a visible pattern for reference. This eliminates
the need for single-particle detectors. Then we rotate the wheels at a very
fast speed $\omega _{f}=\omega _{s}+\Delta \omega $, while putting another
photographic plate on the screen and exposing for the same period of time.
By comparing the current pattern with the initial one, the angle $\Delta
\delta \equiv \delta _{f}-\delta _{s}$ can be measured, where $\delta
_{f}=\omega _{f}a_{2}/v$\ and $\delta _{s}=\omega _{s}a_{2}/v$. Thus the
speed of the response to the change of the boundary condition can be
obtained as%
\begin{equation}
v=\Delta \omega a_{2}/\Delta \delta .  \label{speed v}
\end{equation}

\begin{figure}[tbp]
\includegraphics{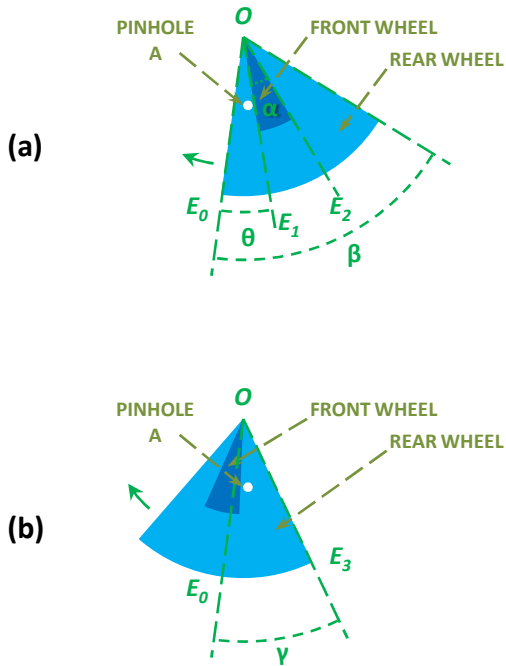}
\caption{The position of the wheels at (a) $t=t_{1}$, and (b)
$t=t_{2}$.} \label{fig:epsart}
\end{figure}

\section{Feasibility and discussions}

In practice, since we would like to determine whether the speed $v$ equals
to the speed $v_{0}$ of the particles in the experiment, it is recommended
to use particles with a non-vanishing mass, so that $v_{0}<c$. Therefore
single-electron interference \cite{ft1,ft12,ft2,ft9} and single cold-atom
interference \cite{ft10,ft8} can both be used. Take for example, consider
the apparatus of the single-electron interference experiment in Ref. \cite%
{ft1}, whose specification was provided in Ref. \cite{ft12}. It used a
convergent electrostatic biprism to take the place of the double-slit. The
distance (denoted as $b$\ in Ref. \cite{ft12}) between the biprism and the
screen (before magnified by the projector lenses) is $a_{2}=6cm$. When the
wire potential of the biprism is $24V$, an interference pattern with a
fringe spacing of $1000\mathring{A}$\ can be obtained (see Fig. 5(e) of Ref.
\cite{ft12}). To implement our experimental scheme, all we need is to add
the wheels shown in Fig. 2 to their apparatus. The wheels should be made of
the same sort of non-magnetic insulating materials used in the biprism so
they will not charge up nor disturb the magnetic field of the lenses. Let
the radius of both the front and rear wheels be $R=10cm$. Since it is
sufficient for us to observe the angle $\delta $\ at only one of the edge of
the sector $OE_{0}E_{3}$, e.g., $OE_{0}$, the sector $OE_{0}E_{3}$ can be
much larger than the size of the entire visible area of the interference
pattern. Therefore the angles $\alpha $\ and $\beta $\ need not to be too
small nor precise so that they can be prepared easily. The hardest part of
the experiment may be that the wheels need to be placed precisely, so that
the tip $E_{0}$ of the rear wheel falls within the visible area of the
interference pattern on the screen when the front wheel starts to cover one
half of the biprism. This is also how the angle $\theta $ between the left
edges of two wheels (as shown in Fig. 3a) is determined. Once this is done,
we rotate the wheels at a low speed, e.g., $\omega _{s}\backsim 10$ round
per second, and turn on the whole system to get a pattern for reference. As
long as the wheels are correctly placed, half of the pattern we observe now
should be identical to the interference pattern without the wheels (i.e.,
Fig. 5(e) of Ref. \cite{ft12}), while the other half should be blurred by
the single-slit diffraction pattern (which should look like a mix of Fig.
5(a) and Fig. 5(e) of Ref. \cite{ft12}). This can also be used as an
approach to verify whether the wheels are placed correctly. After that, we
rotate the wheels at a high speed $\omega _{f}=500$ round per second. Then
as long as the speed we want to measure (i.e., $v$ in Eq. (\ref{speed v}) )
is finite, the position of the dividing boundary between the interference
region and the mixed region in the pattern currently observed should be
different from that of the pattern previously observed at low rotation speed
$\omega _{s}$\ of the wheels. And the difference is most significant on the
location of the screen corresponding to the far end of the rear wheel (the
tip $E_{0}$) when the front wheel starts to cover one half of the biprism.
Rigorously, according to Eq. (\ref{speed v}), the change of the position of
the dividing boundary in the pattern around this location is%
\begin{equation}
x\equiv R\Delta \delta =R\Delta \omega a_{2}/v.
\end{equation}%
The speed of the electron in the experiment in Refs. \cite{ft1,ft12} is $%
v_{0}=1.5\times 10^{8}m/\sec $. Therefore if the speed $v$ equals to $v_{0}$%
, then we get $x=R\Delta \omega a_{2}/v_{0}=1232\mathring{A}$. Even if $v$
equals to the speed $c$ of light, which is the maximum allowed by the theory
of relativity, there is still $x=R\Delta \omega a_{2}/c=616\mathring{A}$.
Both values are comparable to the fringe spacing ($1000\mathring{A}$) thus
are detectable.

On the other hand, if we merely want to determine whether the speed $v$ is
finite or not without caring the relationship between $v$ and $v_{0}$, then
single-photon double-slit interference experiments will be more convenient.
In this case, the distance $a_{2}$ between the double-slit and the screen in
free space can be $\backsim 10^{2}$ times larger than that of the
single-electron interference experiment, and can be made even larger with
optical fibers. Therefore we can observe a significantly larger $x$ and
measure it with higher precision. The visible area of the interference
pattern is also larger, thus the radius of the wheels can be increased too,
so that the speed $\omega _{f}$ can be lower. The disadvantage of these
experiments with photons is that $v_{0}$ is exactly the speed of light. Then
if the experimental result shows that $v$ also equals to the speed of light,
it can hardly provide any information on whether the status of the boundary
condition is learned by the photons themselves or by something else being
exchanged between the experimental instruments.

No matter which type of experiments is used, as long as the result shows
that $v$ is indeed finite, then it seems to support our above interpretation
why our macroscopic world shows classicality though any physical process on
the microscopic scale is quantum. Moreover, if experiments with different
types of particles all prove that $v=v_{0}$, then we can conclude that the
particles learn the boundary condition by themselves. Thus the doubts \cite%
{ft29,ft28} on static experiments for testing Bell's inequality can
be clarified. Also, in double-slit interference type of phenomena,
logically this result can even be understood as an evidence showing
that a single particle indeed passes the two slits simultaneously.
Or if we find $v\neq v_{0}$, then it will suggest that there are
pilot waves or other intriguing mechanism which deliver the
information on the status of the boundary, that travel separately
from the particles.

We thank Prof. Hua-Zhong Li and Prof. Sofia Wechsler for valuable
discussions. The work was supported in part by the NSF of China
under grant Nos. 10975198 and 10605041, the NSF of Guangdong
province under grant No.9151027501000043, and the Foundation of
Zhongshan University Advanced Research Center.

\end{document}